 \documentclass[10pt,twocolumn,twoside,letter]{IEEEtran}

\usepackage{url}
\usepackage{amsmath}
\usepackage{amsfonts}
\usepackage{theorem}
\usepackage[dvips]{epsfig}      
\usepackage{graphicx}
\usepackage{enumerate}
\usepackage{color}                  

\usepackage{verbatim}

\usepackage{amssymb}
\usepackage{amsbsy}
\usepackage{setspace}
\usepackage{url}

\theoremstyle{plain}
\newtheorem{Theorem}{Theorem}

\newtheorem{Proposition}{Proposition}
\newtheorem{Corollary}{Corollary}

 \newcommand{\Ave}{\operatorname{Ave}}
 \newcommand{\Int}{\operatorname{int}}
 \newcommand{\real}{\operatorname{Re}}
 \newcommand{\imag}{\operatorname{Im}}

{\theorembodyfont{\rmfamily} \newtheorem{Remark}{Remark}}
{\theorembodyfont{\rmfamily} }
{\theorembodyfont{\rmfamily} \newtheorem{Example}{Example}}
{\theorembodyfont{\rmfamily} }
{\theorembodyfont{\rmfamily} }

\newcommand {\R}{\mathbb R}

\newcommand{\be}{\begin{equation}}
\newcommand{\ee}{\end{equation}}

\newcommand{\sname}{} \newcommand{\slabel}[1]{\debug{\fbox{\tiny \sname #1}}\label{\sname #1}}
\newcommand{\debug}[1]{}              
\newcommand{\FB}{\begin{figure}[t]\centering} \newcommand{\FE}[2]{\caption{#2 \debug{\fbox{\sname #1}}} \slabel{#1} \end{figure}} \newcommand{\tB}{\begin{table}[hbtp]\centering}
\newcommand{\tE}[2]{\caption{#2 \debug{\fbox{\sname #1}}}\slabel{#1} \end{table}} 

\begin{document}
 \title{Ribosome Flow Model on a Ring\thanks{This research is partially supported by a research grant
  from   the Israeli
Ministry of Science, Technology and Space.
}}


\author{Alon Raveh,  Yoram Zarai,  Michael Margaliot, and Tamir Tuller \IEEEcompsocitemizethanks{\IEEEcompsocthanksitem
A. Raveh is with the School of Electrical Engineering, Tel-Aviv
University, Tel-Aviv 69978, Israel.
E-mail: ravehalon@gmail.com \protect\\
Y. Zarai is with the School of Electrical Engineering, Tel-Aviv
University, Tel-Aviv 69978, Israel.
E-mail: yoramzar@mail.tau.ac.il \protect\\
M. Margaliot is with the School of Electrical Engineering and the Sagol School of Neuroscience, Tel-Aviv
University, Tel-Aviv 69978, Israel.
E-mail: michaelm@eng.tau.ac.il \protect\\
T. Tuller is with the School of Biomedical Engineering and the Sagol School of Neuroscience, Tel-Aviv
University, Tel-Aviv 69978, Israel.
E-mail: tamirtul@post.tau.ac.il \protect\\
}}

\maketitle

\begin{abstract}
The asymmetric simple exclusion process~(ASEP) is an important model from statistical physics
describing  particles that hop randomly from one site to the next
along an ordered  lattice  of  sites, but only  if the next site is empty.
  ASEP has been used to model and analyze numerous multiagent systems with local interactions
  including the flow of  ribosomes along the mRNA strand.

In ASEP with periodic boundary conditions a particle that hops from the last site returns
to the first one. The mean field approximation of this model
  is  referred to as the \emph{ribosome flow model on a ring}~(RFMR).
  The RFMR may be used to model both synthetic and endogenous gene expression regimes.

 We analyze the RFMR  using the theory of monotone dynamical systems.
We show that it admits a continuum of equilibrium points and that
 every trajectory converges to an equilibrium point.
Furthermore, we show that it entrains
to periodic transition rates between the sites.
 We describe the
implications of the analysis results
to  understanding and engineering cyclic mRNA translation in-vitro and in-vivo.

\end{abstract}


\section{Introduction}

The asymmetric simple exclusion process~(ASEP) is an important
  model in statistical mechanics~\cite{TASEP_book}.
ASEP describes   particles that hop along an ordered  lattice  of
  sites.
  The dynamics is stochastic:
  at each time step the particles are scanned,
and every   particle hops to the next site with some probability
if \emph{the next site is empty}.
This  simple exclusion principle allows
modeling  of  the \emph{interaction}
between the particles.
Note that in particular this   prohibits overtaking between
particles.

The term ``asymmetric'' refers to the fact that
there is a preferred direction of movement.
When the movement is unidirectional, some authors use the term
\emph{totally asymmetric simple exclusion process}~(TASEP).
ASEP was first  proposed in 1968~\cite{MacDonald1968}
 as a model for the movement of ribosomes along the mRNA strand
during gene translation. In this context, the lattice models
the mRNA strand, and the particles are the ribosomes.
Simple exclusion corresponds to the fact that
  a ribosome cannot move forward if there is another ribosome
 right in front of it. ASEP has become
  a paradigmatic model for \emph{non-equilibrium} statistical mechanics~\cite{nems2011,solvers_guide}.
  It is used as
 the standard model
 for gene translation~\cite{TASEP_tutorial_2011}, and has also been applied
  to model numerous
 multiagent  systems  with local interactions  including
 traffic flow,  kinesin traffic, the movement
  of ants along a trail,
  pedestrian dynamics
 and ad-hoc communication networks~\cite{TASEP_book,tasep_ad_hoc_nets}.

In ASEP with \emph{open boundary conditions}, the
 lattice boundaries  are open  and the first and last sites
  are connected to two
  external particle reservoirs that drive the asymmetric flow of the particles along the lattice.
   In ASEP with \emph{periodic boundary conditions}  the lattice is closed,
 so that a particle that hops from the last site returns
to the first one.  In particular, the number
of particles on the lattice
is  conserved.

Recently, the mean field approximation of   ASEP with {open boundary} conditions,
called the \emph{ribosome flow model}~(RFM), has been
  analyzed using tools from systems and control theory~\cite{RFM_stability,HRFM_steady_state,RFM_feedback,zarai_infi,RFM_entrain,rfm_concave}.
 In this paper, we
 consider the mean field approximation of ASEP with periodic boundary conditions.
This is a set of~$n$ deterministic nonlinear first-order ordinary
differential equations, where~$n$
is the number of sites, and  each state-variable
describes the occupancy level in one of the sites.
 We refer to this system as the \emph{ribosome flow model on a ring}~(RFMR).
 To the best of our knowledge, this is the first study
  of the RFMR using tools from systems and control theory.

  In the physics literature, many properties have been proven for the
  ASEP with periodic boundary conditions and \emph{homogeneous}  transition rates (see, e.g.~\cite{derr_surv_94}).
  Another case that is amenable to analysis is where the transition rates vary but
	depending on the particles  rather than on the sites (see e.g.~\cite{Ayyer2014} and the references therein). However, an analytical understanding
   of
  ASEP with site-dependent  inhomogeneous transition rates is still pending. In contrast, most of
  the results in this paper  hold for the general case of an inhomogeneous~RFMR.

 The RFMR may model, for example, the translation of a {\em circular}  mRNA or DNA molecule. Circular RNA forms have
 been described in all domains of life (see for example, \cite{Danan2012,Cocquerelle1993,Cell1993,Burd2010,Hensgens1983,Bretscher1968,Bretscher1969}). Specifically, it was shown that in prokaryotes it is possible to regulate translation from circular DNA \cite{Bretscher1968,Bretscher1969}. In the case of eukaryotes, the canonical scanning model requires free ends of the mRNA \cite{Kozak1979}; however, it is well-known   that in these organisms  mRNA is often (temporarily) circularized by translation initiation factors~\cite{Wells1988}.

 We show that the RFMR admits a continuum of equilibrium points, and that
 every trajectory converges to an equilibrium point.
Furthermore, if the transition rates between the sites vary in a  periodic manner,
 with a common period~$T$,
then every trajectory converges to a periodic solution with period~$T$. In other words,
the RFMR \emph{entrains} (or phase-locks) to the periodic excitation. In the particular case where all the transition rates are equal  all the state variables converge to
the same value, namely, the average of all the
 initial values. We discuss the implications of these results to mRNA translation.

The remainder of this paper is organized as follows.  Section~\ref{sec:model} reviews
 the RFMR. Section~\ref{sec:main} details
 the main results. To streamline the presentation, the proofs are placed in  Appendix~A.
  The final section
summarizes  and describes several possible directions for further research.

We use standard notation. For an integer~$i$,~$i_n \in \R^n$ is the~$n$-dimensional column vector with all entries equal
to~$i$.
 For a matrix~$M$,  $M'$ denotes the transpose of~$M$.
We use~$|\cdot|_1:\R^n \to \R_+$ to
denote the~$L_1$ vector norm, that is,~$|z|_1=|z_1|+\dots+|z_n|$.
For a set~$K$, $\Int(K)$ is the interior of~$K$.

\section{The  model}\label{sec:model}
\begin{figure}[t]
  \begin{center}
  \includegraphics[height=14cm]{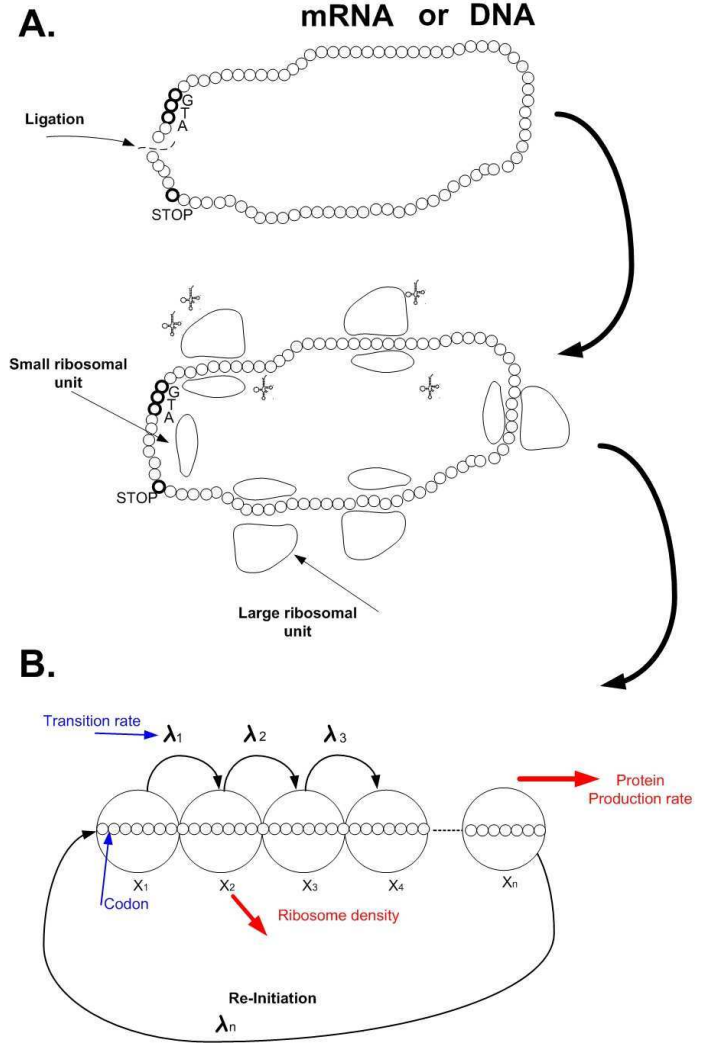}
  \caption{A.~Illustration of  a cyclic  mRNA or DNA  molecule. With this  topology  the ribosomes terminating translation may re-initiate \cite{Skabkin2013,Steege1977,Files1974} translation with high probability. B.~The RFMR as a model for  translation with
	cyclic mRNA. In this context,~$x_i(t)$ is the normalized ribosome density at site~$i$ at time~$t$. The transition rates~$\lambda_i$ depend on factors such as~tRNA abundance.
	}  \label{fig:cmRNA}
  \end{center}
\end{figure}

The \emph{ribosome flow model  on a ring}~(RFMR) is given by
\begin{align}\label{eq:rfm}
                    \dot{x}_1&=\lambda_n x_n (1-x_1) -\lambda_1 x_1(1-x_2), \nonumber \\
                    \dot{x}_2&=\lambda_{1} x_{1} (1-x_{2}) -\lambda_{2} x_{2} (1-x_3) , \nonumber \\
                    \dot{x}_3&=\lambda_{2} x_{ 2} (1-x_{3}) -\lambda_{3} x_{3} (1-x_4) , \nonumber \\
                             &\vdots \nonumber \\
                    \dot{x}_{n-1}&=\lambda_{n-2} x_{n-2} (1-x_{n-1}) -\lambda_{n-1} x_{n-1} (1-x_n), \nonumber \\
                    \dot{x}_n&=\lambda_{n-1}x_{n-1} (1-x_n) -\lambda_n x_n (1-x_1) .
\end{align}
Here~$x_i(t)\in[0,1]$ is the normalized occupancy level  at site~$i$ at time~$t$,   so that~$x_i(t)=0$ [$x_i(t)=1$]
means that site~$i$ is completely empty [full] at time~$t$.
The \emph{transition rates}~$\lambda_1,\dots,\lambda_n$ are
all strictly positive numbers.
To explain this model, consider the equation~$\dot x_2=\lambda_{1} x_{1} (1-x_{2}) -\lambda_{2} x_{2} (1-x_3)$. The term~$r_{12}:=\lambda_{1} x_{1} (1-x_{2})$ represents the flow
of particles   from site~$1$ to site~$2$. This is proportional to the occupancy~$x_1$ at site~$1$
and also to~$1-x_2$, {\em i.e.} the flow decreases as site~$2$ becomes fuller. This is a
relaxed  version of simple exclusion.
The term~$r_{23}:=\lambda_{2} x_{2} (1-x_3)$   represents the flow
of particles from site~$2$ to site~$3$.
The other equations are similar, with the term~$r_{n1}:=\lambda_n x_n (1-x_1)$
appearing both in the equations for~$\dot x_1$ and for~$\dot x_n$ due to the circular structure of the
model (see Fig.~\ref{fig:cmRNA}).

The RFMR encapsulates simple exclusion, unidirectional movement  along the ring,
  and the periodic  boundary condition of~ASEP.
  This is not surprising, as the RFMR is the mean field approximation
of~ASEP with periodic boundary conditions
(see, e.g., \cite[p.~R345]{solvers_guide} and~\cite[p.~1919]{PhysRevE.58.1911}).

Note that we can write~\eqref{eq:rfm} succinctly  as
\[
            \dot{x}_i=\lambda_{i-1}x_{i-1}(1-x_i)-\lambda_i x_i (1-x_{i+1}) ,\quad i=1,\dots,n,
\]
where here and below  every index is interpreted modulo~$n$.
Note also that~$0_n$ [$1_n$] is an equilibrium point  of~\eqref{eq:rfm}.
Indeed, when all the sites are completely free [completely full]
there is no movement of particles between the sites.

Let
\[
           C^n:=\{y \in \R^n: y_i \in [0,1] ,\; i=1,\dots,n\},
\]
i.e.,
the closed unit
cube in~$ \R^n$.
Since the state-variables represent  normalized occupancy  levels,
we always consider initial conditions   $x(0)\in C^n$.
 It is straightforward to verify that~$C^n$ is an invariant set of~\eqref{eq:rfm},
i.e.~$x(0)\in C^n$ implies that~$x(t) \in C^n$ for all~$t\geq 0$.

Note that~\eqref{eq:rfm} implies that
\[
            \sum_{i=0}^n \dot x_i(t)\equiv 0, \text{ for all }t\geq 0,
\]
so  the \emph{total occupancy}~$H(x):=1_n' x$  is conserved:
\be\label{eq:conser}
 H(x(t)) = H(x(0)),\quad \text{for all } t\geq 0 .
\ee
The dynamics thus redistributes the particles between the sites,
but without changing  the total occupancy level. In the context of translation, this means that
the total number of ribosomes on the  mRNA is conserved.

Eq.~\eqref{eq:conser} means that we can reduce
the~$n$-dimensional RFMR  to an~$(n-1)$-dimensional model.
In particular, the RFMR with~$n=2$ can be explicitly solved.
In this case we can obtain explicit expressions for important
quantities, e.g., the rate of convergence to equilibrium.
This solution is detailed in Appendix~B.

If we change the first [last]  equation in~\eqref{eq:rfm} to
$\dot{x}_1 =\lambda_0 (1-x_1) -\lambda_1 x_1(1-x_2)$
[$\dot{x}_n =\lambda_{n-1}x_{n-1} (1-x_n) -\lambda_n x_n  $] we obtain the RFM. This may seem like a minor change, but in fact
the dynamical properties of the RFM and the RFMR are very different.
For example, in  the RFM there is no first integral.
Also,
in the RFM there is a {\sl single} equilibrium point in~$C^n$, whereas as we shall see below
the RFMR has a continuum of equilibrium points in~$C^n$.

  The next section describes several
  theoretical results on the~RFMR. Since the case~$n=2$ is solved in
	Appendix~B, we assume from here on that~$n\geq 3$.
Applications of the analysis  to gene translation are
 discussed  in Section~\ref{sec:discussion}.

 \section{Main results}\label{sec:main}

\subsection{Strong Monotonicity}
A cone~$K\subset \R^n$  defines a partial ordering in~$\R^n$ as follows.
 For two vectors~$a,b \in \R^n$, we write~$a\leq b$ if~$(b-a) \in K$;
  $a<b$ if~$a\leq b$ and~$a \not =b$; and~$a \ll b$ if~$(b -a )\in \Int(K)$.
  The system~$\dot{y}=f(y)$ is called
  \emph{monotone} if~$a \leq b$ implies that~$y(t,a)\leq y(t,b)$ for all~$t \geq 0$.
  In other words, the flow preserves the partial ordering~\cite{hlsmith}.
It is called \emph{strongly monotone} if~$a < b$ implies that~$y(t,a)\ll y(t,b)$ for all~$t > 0$.

From here on we consider
  the  particular case
  where the cone is~$K=\R^n_+$.
  Then~$a\leq b$ if~$a_i\leq b_i$ for all~$i$,
   and~$a \ll b$ if~$a_i <b_i$ for all~$i$.
   A system that is monotone with respect to this partial ordering  is called \emph{cooperative}.


\begin{Proposition}\label{prop:mono}
               Let~$x(t,a)$ denote the solution of the~RFMR at time~$t$ for the
initial  condition~$x(0)=a$.
 For any~$a,b \in C^n$ with~$a \leq b$ we have
                \be\label{eq:abab}
                            x(t,a) \leq x(t,b), \quad \text{for all } t \geq 0.
                \ee
Furthermore, if~$a< b$ then
                \be\label{eq:strongabab}
                            x(t,a) \ll x(t,b), \quad \text{for all } t  > 0.
                \ee
\end{Proposition}
In the context of translation, this means the following.
Consider two possible initial ribosome  densities on the same cyclic
mRNA strand,
$a$ and~$b$ with~$a_i\leq b_i$ for all~$i$, that is, $b$
corresponds to   a higher ribosome density
 at each site. Then the trajectories~$x(t,a)$ and~$x(t,b)$ emanating from these initial conditions
continue to satisfy the same relationship between the densities for
 all time~$t\geq0$.

\subsection{Stability}
Denote the $s$ level set of~$H$ by
\[
L_s:=\{  y \in C^n: 1_n' y= s   \}.
\]
 The next result shows that  every
level set  contains a unique equilibrium  point, and that any
trajectory of the RFMR  emanating  from any point in~$L_s$
  converges to this   equilibrium point.
	In the context of  translation on a cyclic mRNA strand, this means
   that a  perturbations in the distribution of ribosomes along the strand (that does not change
	the total number of ribosomes)
	will not change the asymptotic behavior of the dynamics. It will still converge to the same unique steady
state  ribosome distribution and therefore lead to the same steady-state
 translation rate.
\begin{Theorem} \label{thm:main}
Pick~$s \in [0,n ]$. Then~$L_s$ contains a unique equilibrium point~$e_{L_s}$ of the~RFMR,
and for any~$a \in L_s$,
\[
            \lim_{t\to \infty}x(t,a)=e_{L_s}.
\]
Furthermore, for any~$0\leq s  < p \leq n$, we have
\be\label{eq:linord}
e_{L_s} \ll e_{L_p}.
\ee
\end{Theorem}
 Thm.~\ref{thm:main} implies that the RFMR has a continuum of linearly ordered equilibrium points, namely,
 $\{e_{L_s}: s \in [0,n]\}$, and also that every solution of the RFMR converges to an equilibrium point.

\begin{figure}[t]
  \begin{center}
  \includegraphics[height=7cm]{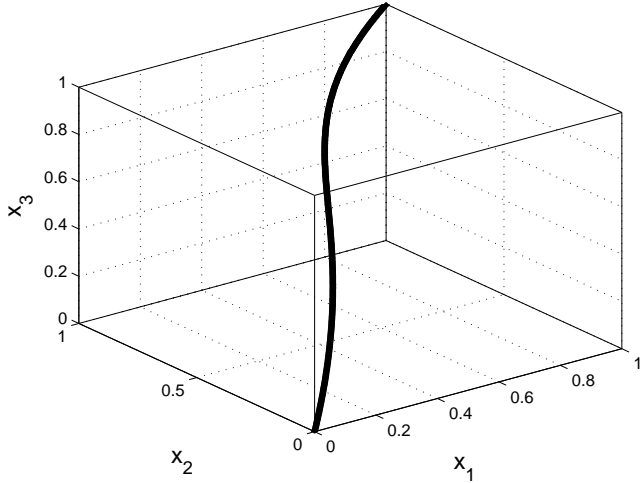}
  \caption{All the equilibrium points of the RFMR  with~$n=3$, $\lambda_1=2$, $\lambda_2=3$, and~$\lambda_3=1$.
  }  \label{fig:alleq}
  \end{center}
\end{figure}

\begin{Example}

Consider the RFMR with~$n=3$, $\lambda_1=2$, $\lambda_2=3$, and~$\lambda_3=1$.
Fig.~\ref{fig:L2} depicts trajectories of this RFMR
  for three initial conditions in~$L_2$:
$[ 1\; 1\; 0 ]'$, $[ 1\; 0\; 1 ]'$, and $[ 0\; 1\; 1 ]'$. It may be observed that
all the trajectories converge to the same equilibrium
point~$e_{L_2}\approx \begin{bmatrix}   0.5380  &  0.6528&    0.8091  \end{bmatrix}'$.
\begin{figure}[t]
  \begin{center}
  \includegraphics[height=7cm]{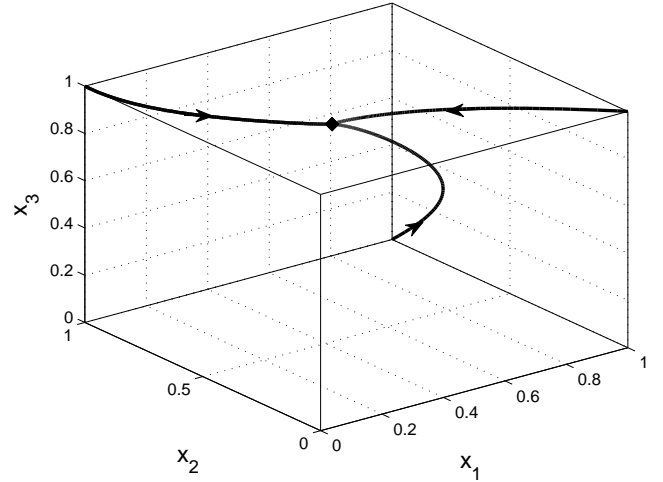}
  \caption{Trajectories of~\eqref{eq:rfm} with~$n=3$
  for three different initial
  conditions in~$L_2$: $[ 1\; 1\; 0 ]'$, $[ 1\; 0\; 1 ]'$, and $[ 0\; 1\; 1 ]'$.
   The equilibrium point~$e_{L_2}$  is marked with a circle. }  \label{fig:L2}
  \end{center}
\end{figure}
Fig.~\ref{fig:alleq} depicts   \emph{all} the equilibrium points of this RFMR.
Since~$\lambda_2>\lambda_1 $ and~$\lambda_2> \lambda_3$, the transition rate into site~$3$ is relatively large.
As may be observed from the figure this leads to~$e_3 \geq e_1$ and~$e_3 \geq e_2$ for every equilibrium point~$e$.~$\square$
\end{Example}

Fix an arbitrary~$s\in [0,n]$.  To simplify the notation, we just write~$e$
instead of~$e_{L_s}$ from here on. Then
\be \label{eqe_L}
                    1_n' e=s,
\ee
and since for~$x=e$ the left-hand side of all the equations
in~\eqref{eq:rfm} is zero,
\begin{align} \label{eq:ep}
                      \lambda_n  e_n (1- {e}_1) & = \lambda_1 {e}_1(1- {e}_2)\nonumber \\&
                      = \lambda_2  {e}_2(1- {e}_3)   \nonumber \\ & \vdots \nonumber \\
                    &= \lambda_{n-1} {e}_{n-1} (1- {e}_n)  .
\end{align}
Thus, the   flow along the chain always converges to a steady-state value
\[
R:=\lambda_i {e}_i(1- {e}_{i+1}),\quad \text {for all } i.
\]

Using~\eqref{eq:ep} and the equation~$1'_n e=s$ it is possible to derive a polynomial equation
for~$e_1$. For example, when~$n=2$
\[
													(\lambda_2-\lambda_1)e_1^2+( \lambda_1(s-1)-
					\lambda_2(s+1) )e_1+\lambda_2 s =0,
\]
whereas for~$n=3$ we get
\begin{align*}
 & \lambda _1 \lambda _3 \left(\lambda _1+\lambda _2+\lambda _3\right) e_1^4 \\
 &+ \left(\lambda _2 \lambda _3^2-\lambda _1^2   \left(\lambda _2+\lambda _3 s\right) -\lambda _3 \lambda _1
   \left(\lambda _2 (2 s-1)+\lambda _3 (s+1)\right)\right)e_1^3 \\
 &     +(\lambda _3 \lambda _1 \left(\lambda _2
   \left(s^2-3\right)+\lambda _3 (2 s-1)\right)\\
   & \;\;+\lambda _1^2 \left(\lambda _2 (s-2)+\lambda _3 (s-1)\right)-\lambda _2 \lambda _3^2
   (s+2) )e_1^2\\
   &+  \lambda _3 \left(\lambda _2 \lambda _3 (2 s+1)+\lambda _1 \left(\lambda _2
   (1-(s-2) s)-\lambda _3 (s-1)\right)\right)e_1\\&-\lambda _2 \lambda _3^2 s=0.
\end{align*}
 These equations can be solved numerically, but their analysis seems non trivial.

\subsection{Differential Analysis}
Differential  analysis is a powerful tool for analyzing nonlinear dynamical systems
(see, e.g.,~\cite{LOHMILLER1998683,entrain2011,sontag_cotraction_tutorial}). The basic idea is to study the difference between
trajectories that emanate from different initial conditions.
The next result shows that the~RFMR is
  \emph{non-expanding} with
  respect to the~$L_1$ norm.
\begin{Proposition}\label{prop:distance}
                                    For any~$a,b \in C^n$,
                                    \be\label{eq:dist_fixed}
                                                |x(t,a)-x(t,b)|_1\leq |a-b|_1,\quad \text{for all } t\geq 0.
                                    \ee
\end{Proposition}
In other words,
 the~$L_1$ distance between trajectories can never increase. From the biophysical point of view, this means that the~$L_1$ difference between two profiles of
ribosome densities, related to two different initial conditions,
   in the same cyclic mRNA/DNA is a non-increasing function of time.

\begin{Example}
Pick~$a,b \in C^n$ such that~$b\leq a$. By monotonicity,~$x(t,b)\leq x(t,a)$ for all~$t\geq 0$,
so~$d(t):=|x(t,a)-x(t,b)|_1=1_n' (x(t,a)-x(t,b))$. Thus,
\begin{align*}
               \dot d(t)&= 1_n'  \dot x(t,a)-1_n' \dot x(t,b)     \\
               &=0-0,
\end{align*}
so clearly in this case~\eqref{eq:dist_fixed} holds with an equality.~$\square$

\end{Example}

Pick~$a \in C^n$, and let~$s:=1_n' a$. Substituting~$b=e_{L_s}$ in~\eqref{eq:dist_fixed}
 yields
\be\label{eq:post}
          |x(t,a)-e_{L_s} |_1\leq |a- e_{L_s} |_1,\quad \text{for all } t\geq 0.
\ee
This means that the convergence to~$e_{L_s}$ is monotone in the sense that the $L_1$ distance to~$e_{L_s}$
can never increase.
Combining~\eqref{eq:post}
with
Theorem~\ref{thm:main}
  implies that every equilibrium point of the~RFMR is \emph{semistable}~\cite{Hui20082375}.

\subsection{Entrainment}

Suppose now that the transition rates along the cyclic mRNA (or DNA)
molecule are periodically time-varying functions of time with a common (minimal)
period~$T>0$.  This may correspond for example to periodically varying abundances of tRNA
due   to the  cell-division cycle that is a periodic program for cell replication.
 A natural question is
will the mRNA densities along the mRNA strand (and thus the translation rate)
 converge to a periodically-varying pattern?

We can study this question using the RFMR as follows.
We say that a function~$f$ is~$T$-periodic if~$f(t+T)=f(t)$ for all~$t$. Assume
that  the~$\lambda_i$s are  time-varying  functions
satisfying:
\begin{itemize}
                        \item there exist~$0<\delta_1<\delta_2$ such that~$\lambda_i(t) \in [\delta_1,\delta_2]$ for all~$t \geq 0$ and all~$i \in \{1,\dots,n\}$.
                        \item there exists a (minimal) $T>0$ such that all the~$\lambda_i$s are~$T$-periodic.
\end{itemize}
We refer to the model in this case as the \emph{periodic ribosome flow model on a ring}~(PRFMR).
\begin{Theorem}  \label{thm:period}
                  Consider the PRFMR.
                  Fix an arbitrary~$s \in [0,n]$. There exists a unique function~$\phi_s:\R_+ \to C^n $, that is~$T$-periodic,
                   and                   \[
                   \lim_{t\to \infty }  |x(t,a)-\phi_s(t) | =0,\quad \text{for all }a \in L_s.
                   \]
\end{Theorem}

In other words, every level set~$L_s$ of~$H$
contains a unique periodic solution, and every
solution of the PRFMR emanating from~$L_s$  converges to this solution.
Thus,  the PRFMR entrains (or phase locks) to the periodic excitation in the~$\lambda_i$s.

 Note that since a constant function is a periodic function for any~$T$,
 Thm.~\ref{thm:period} implies entrainment to a
 periodic trajectory in the particular case where one of the
  $\lambda_i$s oscillates, and all the other are constant.
Note also that Thm.~\ref{thm:main} follows from Thm.~\ref{thm:period}.


\begin{Example}\label{exa:perio}
Consider the RFMR with~$n=3$, $\lambda_1(t) =3$, $\lambda_2(t)=3+2\sin(t+1/2)$, and~$\lambda_3(t)=4-2\cos(2t)$.
 Note that all the~$\lambda_i$s are periodic with a minimal common period~$T=2 \pi $.
Fig.~\ref{fig:period} shows the solution~$x(t,a)$ for~$a=\begin{bmatrix}  0.50& 0.01 & 0.90   \end{bmatrix}'$.
It may be seen that every~$x_i(t)$ converges to a periodic function with period~$2\pi$.~$\square$
\begin{figure}[t]
  \begin{center}
  \includegraphics[height=7cm]{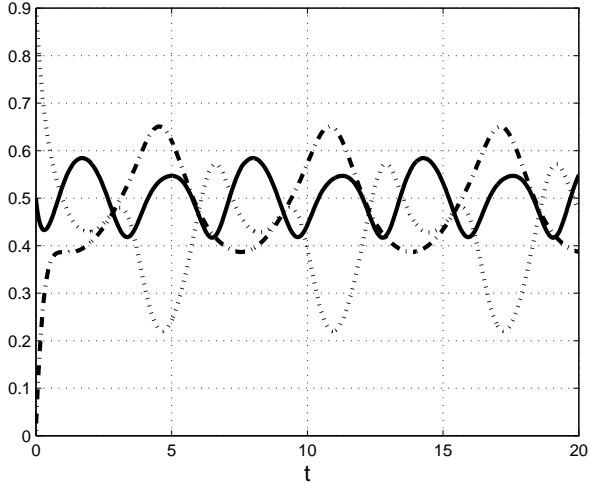}
  \caption{Solution of the PRFMR in Example~\ref{exa:perio}: solid line-$x_1(t,a)$;
  dash-dotted line-$x_2(t,a)$;  dotted line-$x_3(t,a)$.
  }  \label{fig:period}
  \end{center}
\end{figure}
\end{Example}

All the results above hold for general rates~$\lambda_i$. In the particular case where all the rates are equal,
it is possible to provide stronger results.

 \subsection{The homogeneous case} \label{sec:res:hom}
It has been shown experimentally that in some cases
the ribosomal elongation speeds
  along the mRNA sequence are approximately equal~\cite{Ingolia2011}.
To model this case,
assume that
\[
\lambda_1=\dots=\lambda_n:=\lambda_c,
\]
 i.e.,
 all the transition rates are equal, with~$\lambda_c$ denoting
  their common value.
  In this case~\eqref{eq:rfm} becomes:
\begin{align}\label{eq:homo}
                    \dot{x}_1&=\lambda_c x_n (1-x_1) -\lambda_c x_1(1-x_2), \nonumber \\
                    \dot{x}_2&=\lambda_c x_{1} (1-x_{2}) -\lambda_c x_{2} (1-x_3) , \nonumber \\
                             &\vdots \nonumber \\
                    \dot{x}_n&=\lambda_c x_{n-1} (1-x_n) -\lambda_c x_n (1-x_1) .
\end{align}
 We refer to this
   as the \emph{homogeneous ribosome flow model on a ring}~(HRFMR).
Also,~\eqref{eq:ep} becomes
\begin{align}\label{eq:homoep}
                      e_n (1- {e}_1) & = {e}_1(1- {e}_2)\nonumber \\&
                      =   {e}_2(1- {e}_3)   \nonumber \\ & \vdots \nonumber \\
                    &=  {e}_{n-1} (1- {e}_n)  ,
\end{align}
and  it is straightforward to verify that $e=c 1_n$, $c\in \R$,
 satisfies~\eqref{eq:homoep}.

 Define the \emph{averaging operator}~$\Ave(\cdot):\R^n \to \R$   by~$\Ave(z):=\frac{1}{n} 1_n'z$.
\begin{Corollary}\label{coro:hrfm_conv}
 For any~$a \in C^n$ the solution of the HRFMR satisfies
 \be\label{eq:avecon}
            \lim_{t\to\infty}x(t,a)=\Ave(a)1_n.
 \ee
\end{Corollary}
From the biophysical point of view, this means that   equal transition rates along the circular mRNA
lead to convergence to a uniform distribution of the ribosome densities.

Note that~\eqref{eq:avecon} implies that the steady-state flow is~$R=\lambda_c \Ave(a) (1-\Ave(a) )$.
Thus,~$R$ is maximized when~$\Ave(a)=1/2$ and the maximal value is~$R^*=\lambda_c/4$.

\begin{Remark}
It is possible also to give a simple and self-contained proof of Corollary~\ref{coro:hrfm_conv}
using standard  tools
from  the literature on consensus networks~\cite{eger2010}.
Indeed,
pick~$\tau>0$ and let~$i$ be an index such that~$x_i(\tau) \geq x_j(\tau)$ for all~$j \not = i$.
Then
\begin{align*}
                       \dot{x}_i(\tau) &=   x_{i-1}(\tau)(1-x_i(\tau))-  x_i (\tau)(1-x_{i+1}(\tau))\\
                                 &\leq   x_{i }(\tau)(1-x_i(\tau))-  x_i (\tau)(1-x_{i }(\tau))\\
                                 &= 0.
\end{align*}
Furthermore, if~$x_i(\tau) > x_j(\tau)$ for all~$j \not = i$
then~$\dot{x}_i(\tau)<0$.
A similar argument shows that if~$x_i(\tau)\leq x_j(\tau)$ [$x_i(\tau)<x_j(\tau)$] for all~$j \not = i$
then~$\dot{x}_i(\tau)\geq 0$ [$\dot{x}_i(\tau) >  0$].
 Thus, the maximal density never increases, and the minimal density never decreases.
Define~$V(\cdot):\R^n\to\R_+$ by~$V(y):=\max_{i} y_i-\min_i y_i$.
Then~$  V(x(t)) $  strictly
decreases along trajectories of the HRFMR unless~$x(t)  =c 1_n$ for some~$c \in \R$,
and a standard argument (see, e.g.,~\cite{Liu20093122})
implies that the system converges to consensus.
 Combining this with~\eqref{eq:conser} completes the proof of Corollary~\ref{coro:hrfm_conv}.
\end{Remark}

We note in passing that
  Corollary~\ref{coro:hrfm_conv}
implies that the~HRFMR may be interpreted
as a  \emph{nonlinear average consensus network}.
Indeed, every state-variable replaces information with its two nearest neighbors
on the ring only, yet the dynamics
 guarantee that every  state-variable  converges to~$\Ave(a)$.
Consensus networks are recently attracting considerable interest~\cite{eger2010,Consensus07,consen_switch}, and
have many applications in distributed and multi-agent systems.

The physical nature of the underlying model provides a simple explanation for convergence to average consensus.
Indeed, the HRFMR may be interpreted as a  system
  of~$n$ water tanks connected   in a circular topology through  identical  pipes.
The flow in this system is driven by the imbalance in the water levels,
 and the state always converges to a homogeneous distribution of water in the tanks.
 Since the system is closed, this corresponds to average consensus.

We next analyze  the linearized model of the HRFMR near an equilibrium point
to obtain  information on the convergence rate and the amplitude of
the oscillations.

\subsubsection{Local analysis}
 \label{sec:res:local}

Every trajectory
 of the HRFMR converges to~$c 1_n$, where~$c$ depends on the initial condition.
Let~$y:=x-c 1_n$.
 Then a calculation shows that the linearized dynamics of~$y$  is given by
\be\label{eq:ydotisQy}
\dot y= Q y,
\ee
 where
\begin{equation}
            Q:=\begin{bmatrix}
            -1& c & 0 & 0&\dots &0 & 1-c \\
            1-c& -1 & c & 0& \dots& 0 & 0 \\
            0& 1-c & -1 & c& \dots& 0 & 0 \\
            \vdots\\
            c& 0 & 0 & 0& \dots &1-c & -1 \\
                \end{bmatrix}.
\end{equation}
By known-results on   circulant  matrices  (see, e.g.,~\cite{mat_ana_sec_ed}),
  the eigenvalues of~$Q$ are
\be\label{eq:eigQ}
         \gamma_\ell=-1+c w^{\ell-1}+(1-c)w ^{(\ell-1)(n-1)}                    ,\quad \ell=1\dots, n,
\ee
where~$w :=\exp(2 \pi  \sqrt{-1}  /n)$,
and the corresponding eigenvectors are
\begin{equation}\label{lineigvec}
  v^\ell := \begin{bmatrix} 1 & \omega^{(\ell-1)}   & \dots & \omega^{(\ell-1)(n-1)} \end{bmatrix}'  ,\quad \ell=1\dots, n.
\end{equation}
In particular,~$\gamma_1=0$, with corresponding eigenvector~$v^1=1_n$.
This
is a consequence  of the continuum of equilibria in the HRFMR.

Note that
\begin{align}\label{eq:realval}
\nonumber \real(\gamma_\ell) &=    -1+ \cos(2\pi(\ell-1)(n-1)/n)\\ \nonumber &+c ( \cos(2\pi(\ell-1) /n)-\cos(2\pi(\ell-1)(n-1)/n)   )\nonumber \\
&= -1+\cos(2\pi(\ell-1)/n),
\end{align}
and this implies that
\begin{align*}
\real(\gamma_\ell)\leq     \real(\gamma_2)= \cos(2\pi/n)-1, \quad \ell=2,\dots,n.
\end{align*}
Thus, for~$x(0)$ in the vicinity of the equilibrium
\be\label{eq:ffgg}
            |x(t)-c 1_n|\leq\exp(    ( \cos(2\pi /n) - 1 )   t )|x(0)-c 1_n|.
\ee
 The exponential convergence rate decreases with~$n$. For example, for~$n=2$,
$ \cos(2\pi /n) -1= -2$, whereas for~$n=10$,  $ \cos(2\pi/n) -1\approx -0.191$.
In other words, as the length of the chain increases the
convergence rate decreases. This is the price paid for
the fact that each site
``communicates'' directly  with its two neighboring sites only.

\begin{figure}[t]
  \begin{center}
  \includegraphics[height=7cm]{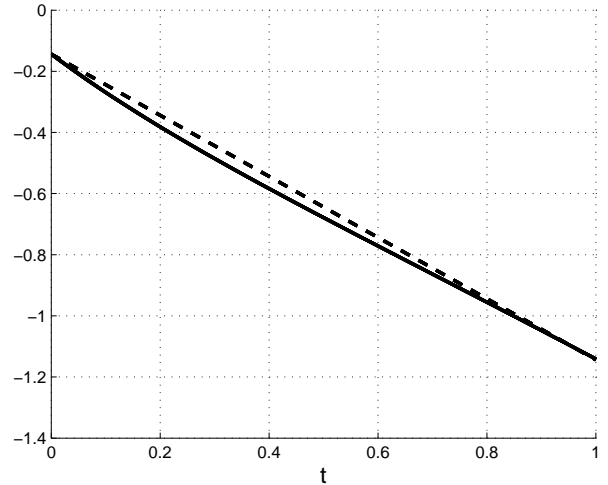}
  \caption{ $\log(|x(t )- (1/4) 1_4|)$ in the HRFMR
   with~$n=4$ and~$x(0)=[1\;0\;0\;0]'$ as a function of~$t$
   (solid line). Also shown is the function~$  -t+ \log( \sqrt{3/4})  $
   that is obtained from the local analysis (dashed line).   }  \label{fig:cr}
  \end{center}
\end{figure}

Our simulations suggest that~\eqref{eq:ffgg}
actually provides a reasonable approximation for the real convergence rate (i.e.,
not only in the vicinity of the equilibrium point) in the HRFMR. The next example demonstrates this.
\begin{Example}
                      Consider the HRFMR with~$n=4$.
                     Fig.~\ref{fig:cr} depicts~$\log(|x(t )- (1/4) 1_4|)$ for the initial
                     condition~$x(0)=\begin{bmatrix} 1& 0 & 0 &0 \end{bmatrix}'$.
                     Note that here~$\log(|x(0 )- (1/4) 1_4|)=\log(\sqrt{3/4})$.
                      In this case,~$\real(\gamma_2)=-1$,
                      so~\eqref{eq:ffgg} becomes
                       $\log( |x(t)-c 1_n| ) \approx -t+ \log(|x(0)-c 1_n|)$.
                     Also shown is the graph of~$-t+ \log( \sqrt{3/4}) $.
                     It may be seen that the real convergence rate is slightly faster than
                     the estimate~\eqref{eq:ffgg}.~$\square$
\end{Example}

Eq.~\eqref{eq:realval} implies that~$c$ does not affect the convergence rate in the linearized system.
It does however affect the oscillatory behavior until convergence.
To see this, note that   the solution of~\eqref{eq:ydotisQy} is
\be\label{eq:soly}
            y(t)=\sum_{i=1}^n s_i \exp(\gamma_i t)v^i,
\ee
where the~$s_i$s satisfy
\[
                y(0)=\sum_{i=1}^n s_i  v^i.
\]
Since~$\gamma_1=0$ and~$\real(\gamma_i)<0$  for all~$i>1$,~\eqref{eq:soly}
 implies that~$\lim_{t\to\infty}y(t)=s_1 v^1$, so~$s_1=0$.
The three eigenvalues with the largest real part are~$\gamma_1,\gamma_2,$ and~$\gamma_n$, so for
large~$t$,
\begin{align*}
                    y(t)\approx s_2 \exp(\gamma_2 t)v^2+s_n \exp(\gamma_n t)v^n.
 \end{align*}
 It follows from~\eqref{eq:eigQ} that
 \begin{align*}
 \alpha:=\real(\gamma_2)&=\real (\gamma_n) =  \cos(2\pi/n)-1,\\
 \beta:=\imag(\gamma_2)&=-\imag (\gamma_n) =(2c-1) \sin(2\pi/n),
\end{align*}
so~$v^n=\bar {v}^2$,  $s_n=\bar {s}_2$
and this yields
\begin{align*}
  y  (t) & \approx 2 |p|  \exp( \alpha  t ) \cos \left( \beta t +\angle p \right )
							    ,
\end{align*}
where~$p:=s_2 v^2$.
   The oscillatory behavior
   thus depends on~$c$.
For~$c=1/2$ the solution has no oscillations at all,
and as~$c$ moves away from~$1/2$
  the oscillations  become larger.
The reason  for this is that for~$c=1/2$ the linear equation~\eqref{eq:ydotisQy} corresponds to the
case where each agent weighs the contribution from its two neighbors  equally.
As~$c$ moves away from~$1/2$ the weights become different and this
imbalance leads to oscillations until the state-variables converge to the correct values.

\begin{figure}[t]
  \begin{center}
  \includegraphics[height=7cm]{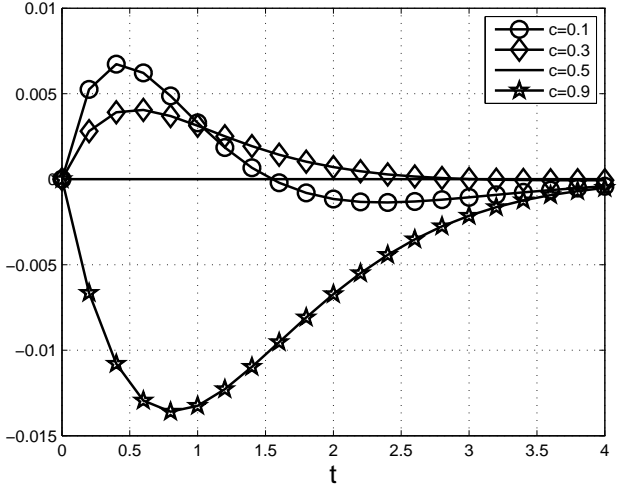}
  \caption{ The function~$d(t,c)$ for several  values of~$c$.    }  \label{fig:osci}
  \end{center}
\end{figure}

\begin{Example}
Let~$z(c):=\begin{bmatrix}  c+0.1 & c-0.1& c  \end{bmatrix}'$.
We simulated the trajectories of the HRFMR with~$n=3$ for four initial conditions:
$x_0=z(c)$, with~$c=0.1,0.3,0.5$, and~$0.9$.
Note that~$\frac{1}{3}1_3'z(c) =c$,
so~$\lim_{t\to\infty }x(t,z(c))=c1_3$.
Fig.~\ref{fig:osci}
depicts
\[
d(t,c):= (x_1(t,z(c))-c )-(x_1(t,z(1/2))-1/2)
\]
 as a function of~$t$. It may be seen that as~$|c-1/2|$
 increases the oscillations increase. This agrees with the analysis above.
\end{Example}

\subsection{The Case of a Single Slow Rate}
It is interesting to consider the case where all the transition rates are equal to~$\lambda_c$, except for
$\lambda_1$ that has value~$\lambda_q$, with~$\lambda_q<\lambda_c$. For TASEP with periodic boundary conditions
this case has been studied in~\cite{finite_size_effect, exact_asep_blockage}.
 It corresponds to  a translation
regime with  the   elongation rates  basically uniform (with rate~$\lambda_c$),
 yet the re-initiation, and possibly termination, rate
 (modeled by~$\lambda_q$) is   slower.
In this case, Eq.~\eqref{eq:ep} of the RFMR  becomes
\begin{align} \label{eq:ep_with_block}
                      e_n (1- {e}_1) & =\frac{ \lambda_q}{\lambda_c} {e}_1(1- {e}_2)\nonumber \\&
                      =   {e}_2(1- {e}_3)   \nonumber \\ & \vdots \nonumber \\
                    &=  {e}_{n-1} (1- {e}_n)  ,
\end{align}
so we  assume from here on, without loss of generality, that~$\lambda_c=1$.

\begin{figure}[t]
  \begin{center}
  \includegraphics[height=7cm]{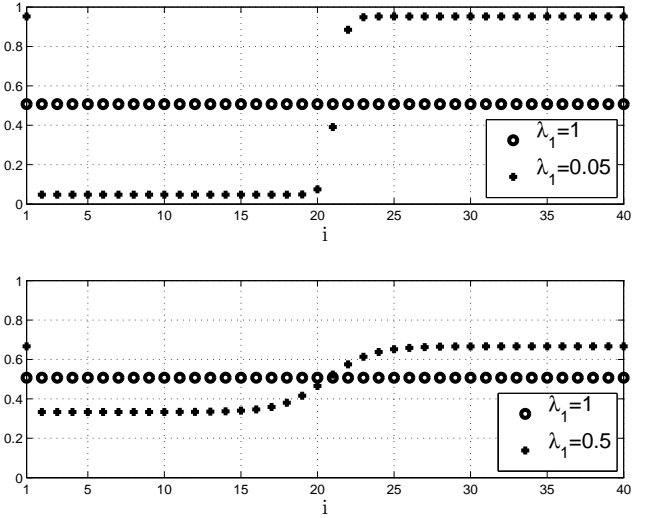}
  \caption{Steady-state occupancy levels $e_i$, $i=1,\dots,40$, in  a
	RFMR with $n=40$, $s=20.3$ and $\lambda_2=\dots=\lambda_n=1$.
	Upper figure: $\lambda_1=0.05$ ('+') and $\lambda_1=1$ ('o') (i.e., the HRFMR).
	Lower figure: $\lambda_1=0.5$ ('+') and $\lambda_1=1$ ('o'). }
		\label{fig:n40_lc_lq}
  \end{center}
\end{figure}

When~$ |\frac{s}{n}-\frac{1}{2}| $ is small, i.e., the normalized total occupancy is close to~$1/2$
 the steady-state~$e$ has the form depicted in Fig.~\ref{fig:n40_lc_lq}.
The slow rate yields an
 increase  [decrease]  in the steady-state particle density to the immediate
left [right]. In other words, the slow rate induces a ``traffic jam''
segregating  the steady-states
  into high- and low-density regions.
Near the  slow site the densities become  more or less uniform
with a low density~$e_l$ and a high density~$e_h$.
Substituting this in~\eqref{eq:ep_with_block}
gives
\begin{align*} \label{eq:ep_with_block}
                      e_h (1- {e}_h) & \approx { \lambda_q}  {e}_h (1- {e}_l)\nonumber \\&
                      \approx    {e}_l(1- {e}_l) ,
\end{align*}
so $e_l\approx  \frac{\lambda_q}{1+\lambda_q }$, and $e_h \approx
 \frac{ 1}{1+\lambda_q }$.
Note that this implies that~$e_l+e_h \approx 1$.
For example, for~$\lambda_q=0.05$ this gives~$e_l\approx 0.047619$, $e_h\approx 0.952381$,
 and this agrees well with
the case depicted in  Fig.~\ref{fig:n40_lc_lq}.

The steady-state flow is thus
\begin{align*}
			R&      =  e_h (1- {e}_h)      \\ &\approx \frac {  \lambda_q} {(1+\lambda_q)^2}.
\end{align*}

Let~$m_l$ [$m_h$] denote the number of~$e_i$s satisfying~$e_i\approx e_l$,
so that~$m_h:= n-m_l$  approximates the number of~$e_i$s satisfying~$e_i\approx e_h$.
Then the equation~$s\approx m_l e_l+m_h e_h$ yields
\[
						m_l\approx \frac{   n-s(1+\lambda_q)  } {1-\lambda_q}.
\]
For example, for the case~$n=40$, $s=20.3$, $\lambda_q=0.05$ this gives~$m_l\approx 19.6684$, and this agrees well with
the case depicted in  Fig.~\ref{fig:n40_lc_lq}.

 From the biophysical point of view, these results
suggest that  a slower re-initiation step (that may interact and delay the termination step) will lead
to a ribosomal ``traffic jam'' before the STOP codon.

\section{Discussion}
 \label{sec:discussion}

The ribosome flow model on a ring~(RFMR)
is the mean field approximation of ASEP with periodic boundary conditions.
 We analyzed the RFMR using tools from  monotone dynamical systems theory.
 Our results show that the
  RFMR has several nice properties.  It is an irreducible
   cooperative dynamical system admitting a continuum of linearly ordered
equilibrium points, and
every trajectory converges to an equilibrium point. The RFMR
 is on the ``verge of contraction'',
and it entrains
  to periodic transition rates.

Topics for further research include the following.    ASEP
with periodic boundary conditions has been studied extensively in the physics literature
and many explicit results are known. For example, the time
scale until  the system
  relaxes to the (stochastic) steady state is known~\cite{solvers_guide}.
  A natural research direction is
based on extending such results to the RFMR.

For the RFM, that is, the mean-field approximation of ASEP with \emph{open}
boundary conditions, it has been shown that the steady-state translation rate~$R$ satisfies the equation
\[
            0=f(R),
\]
where~$f$ is a continued fraction~\cite{RFM_stability}. Using the well-known relationship between
continued fractions and tridiagonal matrices (see, e.g.,~\cite{wall_contin_frac}) yields
that~$R^{-1/2}$ is the Perron root of a certain non-negative symmetric
tridiagonal matrix
with entries that depend on the~$\lambda_i$s~\cite{rfm_concave}.
This has many applications. For example it implies that~$R=R(\lambda_0,\dots,\lambda_n)$
in the RFM is a strictly concave function on~$\R^{n+1}_+$~\cite{rfm_concave}.
 It also implies that
sensitivity analysis in the RFM is   an eigenvalue sensitivity problem~\cite{RFM_SENSITIVITY}.
An interesting research 
 question is whether~$R$ in the RFMR can also be described using such equations.

Another interesting topic for further research is studying a network of several connected RFMs.
The output of each~RFM is divided between the inputs of all the RFMs. This models competition for the
ribosomes in  the cell.   Assuming that the     system is closed then
leads to network of interconnected~RFMRs.

The RFMR may be  used in the future for analyzing novel synthetic circular DNA or mRNA molecules for protein translation in-vitro or in-vivo. Such devices have potential advantages with respect to linear mRNA,
 since they do not include free ends and may thus be more stable.
 For example, in {\em E. coli}  RNA degradation often begins with conversion of the $5 '$-terminal triphosphate to a monophosphate, creating a better substrate for internal cleavage by RNase E \cite{Richards2011}. In eukaryotes, such as {\em Saccharomyces cerevisiae}, intrinsic mRNA decay initiate with deadenylation that causes the shortening of the poly(A) tail at the $3 '$ end of the mRNA, followed by the removal of the cap at the $5 '$ end by the decapping enzyme, which leads to a rapid $5 ' \rightarrow 3 '$ degradation of the mRNA by an exoribonuclease \cite{Shirley1998}.
In eukaryotes, the canonical scanning model requires free ends of the mRNA~\cite{Kozak1979}.
However,
it may be possible to design circular DNA or mRNA molecules by using
  \emph{Internal Ribosome Entry Sites} (IRESes)~\cite{Hellen2001}. Such an experimental system (or a similar system) can be used in the future for evaluation the theoretical results reported in this study.



\section*{Appendix A: Proofs}
 {\sl Proof of Prop.~\ref{prop:mono}.}
Write the RFMR~\eqref{eq:rfm} as~$\dot x=f(x)$. The
  Jacobian matrix~$J(x):=\frac{\partial f}{\partial x}(x)$ is given in~\eqref{eq:jacobian}.
   This matrix has nonnegative off-diagonal entries for all~$x \in C^n$.
Thus, the~RFMR is a  {cooperative system}~\cite{hlsmith}, and this implies~\eqref{eq:abab}.
 Furthermore, it is straightforward to verify that~$J(x)$ is an irreducible matrix for all~$x \in \Int(C^n)$,
 and this implies~\eqref{eq:strongabab} (see, e.g.,~\cite[Ch.~4]{hlsmith}).~\IEEEQED

\begin{figure*}[!t]
\normalsize
\begingroup\makeatletter\def\f@size{7}\check@mathfonts
\begin{equation}\label{eq:jacobian}
 J(x)=\begin{bmatrix}
-\lambda_n x_n - \lambda_1 (1-x_2) & \lambda_1 x_1                                 & 0            &           &  0&     \lambda_n (1-x_1)\\
\lambda_1(1-x_2)                   & -\lambda_1 x_1-\lambda_2(1-x_3)               & \lambda_2 x_2     &        & 0   & 0\\
0                                  &  \lambda_2  (1-x_3)                           & -\lambda_2 x_2 -\lambda_3(1-x_4)  & \dots & 0 &0\\
                                    &                                              &     \vdots \\
0                                   &                                             0 &      0          &                  & -\lambda_{n-2}x_{n-2} -\lambda_{n-1} (1-x_n)  & \lambda_{n-1}x_{n-1} \\
\lambda_n x_n                                       &                   0 & 0                  &            & \lambda_{n-1}(1-x_n) & -\lambda_{n-1}x_{n-1} -\lambda_n(1-x_1)
\end{bmatrix}
\end{equation}
\endgroup
\hrulefill
\vspace*{4pt}
\end{figure*}

 {\sl Proof of Thm.~\ref{thm:main}.}
Since the RFMR is  a cooperative irreducible system with~$H(x)=1_n' x$ as a first integral,
    Thm.~\ref{thm:main} follows from the results in~\cite{mono_plus_int} (see also~\cite{Mierc1991}
    and~\cite{mono_chem_2007}
 for some related ideas).~\IEEEQED

  {\sl Proof of Prop.~\ref{prop:distance}.}
Recall that the matrix measure~$\mu_1(\cdot):\R^{n \times n} \to \R$ induced by the~$L_1$ norm
is
\[
            \mu_1(A)=\max \{c_1(A),\dots, c_n(A)\},
\]
where~$c_i(A):=a_{ii}+\sum_{k\not = i} | a_{ki} | $,
i.e.   the sum of entries in column~$i$ of~$A$, with the off-diagonal entries taken with absolute value~\cite{vid}.
For the Jacobian of the RFMR, we have~$c_i(J(x))=0$  for all~$i$ and all~$x \in C^n$, so~$\mu_1(J(x))=0$.
Now~\eqref{eq:dist_fixed} follows from standard results in contraction theory (see, e.g.,~\cite{entrain2011}).~\IEEEQED

{\sl Proof of Thm.~\ref{thm:period}.}
Write the PRFMR as~$\dot x= f(t,x)$. Then~$f(t,y)=f(t+T,y)$ for all~$t$ and~$y$.
Furthermore,~$H(x)=1_n' x$ is a first integral of the PRFMR.
Now Thm.~\ref{thm:period} follows from the results in~\cite{mono_periodic} (see also~\cite{mono_periodic_96}).~\IEEEQED


  {\sl Proof of Corollary~\ref{coro:hrfm_conv}.}
Let~$s:=1_n'a$. Then~$L_s$ contains~$\Ave(a)1_n$ and this  is an equilibrium point.
The proof now follows immediately from Thm.~\ref{thm:main}.~\IEEEQED

\section*{Appendix B: Solution of the RFMR with~$n=2$}
                Consider~\eqref{eq:rfm} with~$n=2$, i.e.
\begin{align}\label{eq:two_dim}
                    \dot{x}_1&=\lambda_2 x_2 (1-x_1) -\lambda_1 x_1(1-x_2), \nonumber \\
                    \dot{x}_2&=\lambda_{1}x_{1} (1-x_2) -\lambda_2 x_2 (1-x_1) .
\end{align}
                We assume that~$x(0) \not =0_2$ and~$x(0)\not =1_2$, as these are equilibrium points of the
                dynamics.
                Let~$s:=x_1(0)+x_2(0)$.
                Substituting~$x_2(t)=s-x_1(t)$   in~\eqref{eq:two_dim} yields
 \begin{align}\label{eq:ricc}
                    \dot{x}_1&=\lambda_2 (s-x_1) (1-x_1) -\lambda_1 x_1(1-s+x_1)  \nonumber\\
                             &=\alpha_2 x_1^2 +\alpha_1 x_1+\alpha_0,
\end{align}
where
\begin{align*}
\alpha _{2}&:=\lambda_2-\lambda_1,\\
\alpha_1&:=(\lambda_1-\lambda_2)s  - \lambda_1 - \lambda_2 ,\\
\alpha_0&:=s \lambda_2.
 \end{align*}
  If~$\lambda_1=\lambda_2$ then~\eqref{eq:ricc} is a linear differential
  equation and its solution is
\begin{align} \label{eq:n2simple}
x_1(t) =
                  \frac{s}{2}(1-\exp(-2\lambda_1 t))  +  x_1(0)   \exp(-2\lambda_1 t),
\end{align}
 so
 \begin{align}\label{eq:x2exp}
 x_2(t)&=s-x_1(t)\nonumber \\
 &=\frac{s}{2}(1 + \exp(-2\lambda_1 t))  -  x_1(0)   \exp(-2\lambda_1 t) \nonumber\\
 &=\frac{s}{2}(1 - \exp(-2\lambda_1 t))  +  x_2(0)   \exp(-2\lambda_1 t).
 \end{align}
 In particular,
 \be \label{eq:xcpns/2}
 \lim_{t\to\infty}x(t)= (s/2)1_2,
 \ee
{\em i.e.}, the state-variables converge at an exponential rate
to the average of their initial values.

In the context of translation on  a circular mRNA/DNA this means that  uniform translation rates
are expected to yield    a steady-state of
   uniform ribosomal distributions  along the transcript, and that the convergence to this steady-state is fast.

  If~$\lambda_1 \not = \lambda_2$ then~\eqref{eq:ricc} is a   Riccati  equation (see, e.g.~\cite{riccati2007}),
whose solution is
\be\label{eq:sol2}
                x_1(t)= \frac {  -\alpha_1 - \sqrt{\Delta}  \coth( \sqrt{\Delta}(t-t_0)/2 ) } {2 \alpha_2},
\ee
   where
   \begin{align*}
   \Delta & := \alpha_1^2-4 \alpha_2 \alpha_0= (s-1)^2 (\lambda_1 - \lambda_2)^2  + 4 \lambda_1 \lambda_2   ,\\
  t_0&:=\frac{2}{\sqrt{\Delta}}
   \coth^{-1}\left( \frac{2x_1(0)\alpha_2+\alpha_1}{ \sqrt{\Delta} } \right).
  \end{align*}
  Note that    since the~$\lambda_i$s are positive, $\Delta>0$. Also, a
 straightforward calculation  shows that~$t_0$ is well-defined
  for all~$x_1(0) \in [0,1]$.
Note that~\eqref{eq:sol2} implies that
\[
 \lim_{t\to \infty} x(t)=\frac{1}{2\alpha_2}  \begin{bmatrix}
 {  -\alpha_1 - \sqrt{\Delta}   }
& {  2 \alpha_2 s+ \alpha_1 + \sqrt{\Delta}   }
   \end{bmatrix}'.
   \]
The identity
\be\label{eq:coth_idnt}
                    \coth\left ( \frac{t}{2}\sqrt{\Delta} \right )-1=\frac{2}{\exp(\sqrt{\Delta}  t)-1}
\ee
   implies that for
     sufficiently large values of~$t$ the convergence is  with   rate~$\exp(-\sqrt{\Delta}   t)$.
Thus, the convergence rate depends on~$\lambda_1$, $\lambda_2$, and~$s$.

\begin{figure}[t]
  \begin{center}
  \includegraphics[height=7cm]{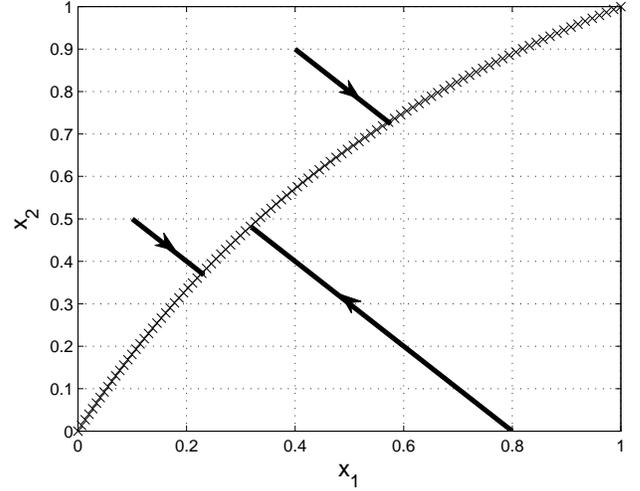}
  \caption{Trajectories of~\eqref{eq:rfm} with~$n=2$,~$\lambda_1=2$ and~$\lambda_2=1$ for three initial conditions.
  The dynamics admits a continuum of equilibrium points  marked by~$+$.    }  \label{fig:dyn2}
  \end{center}
\end{figure}

 Summarizing, when~$n=2$ every  trajectory of the RFMR follows the straight line from~$x(0)$ to an
 equilibrium point~$e=e(\lambda_1,\lambda_2,s)$.
  In particular, if~$a,b \in C^2$ satisfy~$1_2 'a=1_2'b$
  then the solutions emanating from~$a$ and from~$b$ converge to the same
  equilibrium point. Fig.~\ref{fig:dyn2} depicts the trajectories of the RFMR with~$n=2$,
$\lambda_1=2$ and~$\lambda_2=1$ for three initial conditions.

To study entrainment in this case,
                consider the RFMR with~$n=2$,~$\lambda_1(t)=3q(t)/2$,
                and~$\lambda_2(t)=q(t)/2$,
               where~$q(t)$ is a strictly positive and periodic function. Then~\eqref{eq:ricc} becomes
 \be\label{eq:perq}
  \dot{x}_1 =  ( -x_1^2 + (s-2) x_1+s/2 ) q.
 \ee
 Assume that
 \be\label{eq:onotc}
                            x_1^2(0)<s/2.
 \ee
 It is straightforward to verify that in this case
 the solution of~\eqref{eq:perq} is
 \[
            x_1(t)= (s/2)-1  + z \tanh \left(k+z \int_0^t q(s)ds \right ),
 \]
 where
 \[
 z:=\frac{\sqrt{ 3+(s-1)^2 }}{2},
 \]
 and
 \[
 k:=\tanh^{-1}\left( ( x_1(0)+1-s/2  )/z\right ).
 \]
 Note that~\eqref{eq:onotc} implies that~$k$ is well-defined.
Suppose, for example, that~$q(t)=2+\sin(t)$.
Then~$\lambda_1(t)$ and~$\lambda_2(t)$ are periodic with period~$T=2\pi$.
In this case,
 \[
            x_1(t)= (s/2)-1  + z \tanh \left(k+z ( 2t+1-\cos(t)  ) \right ),
 \]
 and~
 \begin{align*}
 x_2(t)&=s-x_1(t)\\&=  (s/2) +1- z \tanh \left(k+z ( 2t+1-\cos(t)  ) \right )  .
 \end{align*}
    Thus, for every~$a \in L_s$, ~$\lim_{t \to \infty} x(t,a) = \phi_s(t)$,
     where $\phi_s(t) \equiv \begin{bmatrix}
     (s/2)-1  + z & (s/2)+1-z  \end{bmatrix}'$ (which is of course periodic with period~$T$).


\end{document}